\newcommand{\be}{\begin{equation}}\newcommand{\ee}{\end{equation}}
\newcommand{\bea}{\begin{eqnarray}}\newcommand{\eea}{\end{eqnarray}}
\newcommand{\nn}{\nonumber \\}
\newcommand{\hK}{{\hat K}}
\newcommand{\hD}{{\hat D}}
\newcommand{\cD}{{\cal D}}
\newcommand{\p}[1]{(\ref{#1})}

\documentstyle[12pt]{article}

\begin{document}
\setcounter{page}0
\renewcommand{\thefootnote}{\fnsymbol{footnote}}
\thispagestyle{empty}
\begin{flushright}
hep-th/0206126\vspace{2cm}\\ \end{flushright}
\begin{center}
{\large\bf AdS/CFT EQUIVALENCE TRANSFORMATION}
\vspace{0.5cm} \\
S. Bellucci\footnote{bellucci@lnf.infn.it } \vspace{0.5cm} \\
{\it INFN-Laboratori Nazionali di Frascati, \\
 C.P. 13, 00044 Frascati, Italy \vspace{0.5cm} }\\
E. Ivanov\footnote{eivanov@thsun1.jinr.ru }
and S. Krivonos\footnote{krivonos@thsun1.jinr.ru }
  \vspace{0.5cm} \\
{\it Bogoliubov  Laboratory of Theoretical Physics, JINR,\\
141980 Dubna, Moscow Region, Russia} \vspace{1.5cm} \\
{\bf Abstract}
\end{center}

\noindent We show that any conformal field theory in $d$-dimensional Minkowski space,
in a phase with spontaneously broken conformal symmetry and with the dilaton
among its fields, can be rewritten in terms of the static gauge $(d-1)$-brane on AdS$_{(d+1)}$
by means of an invertible change of variables. This nonlinear holographic
transformation maps the Minkowski space coordinates
onto the brane worldvolume ones and the dilaton onto the transverse AdS brane coordinate. One of the
consequences of the existence of this map is that any $(d-1)$-brane worldvolume
action on AdS$_{(d+1)}\times X^m$ (with $X^m$ standing for the sphere $S^m$ or
more complicated curved manifold)
admits an equivalent description in Minkowski space as a nonlinear and higher-derivative
extension of some conventional conformal field theory action, with the conformal group being realized
in a standard way. The holographic transformation explicitly relates the standard
realization of the conformal group to its field-dependent nonlinear realization as the isometry group of
the brane AdS$_{(d+1)}$ background. Some possible implications of this transformation,
in particular, for the study of the quantum effective action of ${\cal N}=4$ super Yang-Mills
theory in the context of AdS/CFT correspondence, are briefly discussed.

\newpage
\renewcommand{\thefootnote}{\arabic{footnote}}
\setcounter{footnote}0
\setcounter{equation}0
\section{Introduction}
The cornerstone of AdS/CFT correspondence \cite{mald,gkp,wit,rev} is the hypothesis
that the isometry group of an AdS$_n\times S^m$ background in which some type IIB
string theory and related supergravity live is identical to the standard conformal group
(times the group of internal $R$ symmetry) of the appropriate conformal field
theory defined on the $(n-1)$-dimensional Minkowski space considered as a boundary of AdS$_n$.
The full supersymmetric version of this correspondence deals with the bulk and
boundary realizations of superconformal groups including conformal and $R$-symmetry
groups as bosonic subgroups.

It was shown in \cite{mald}, \cite{5}-\cite{nscm} that the invariance group of the worldvolume action
of some probe brane in an AdS$_n\times S^m$ background (e.g., a D3-brane in AdS$_5\times S^5$)
can be realized as a field-dependent modification
of the standard (super)conformal transformations of the worldvolume. In \cite{solv} it was
demonstrated that such a realization of the AdS isometry corresponds to the
choice of the special `solvable subgroup' parametrization of the AdS background. In the spirit of
the AdS/CFT correspondence (and some other hypotheses of similar nature), the AdS
superbrane worldvolume actions are expected to appear as the result of summing up
leading and subleading terms in the low-energy
quantum effective actions of the corresponding Minkowski space (super)conformal field
theories in the phase with spontaneously broken (super)conformal symmetry
(e.g., the AdS$_5\times S^5$ D3-brane action \cite{D3} and its some modifications
should be recovered in this way from the effective action of ${\cal N}=4$ SYM theory
in the Coulomb branch \cite{chep,mald2,mald,ats}).
In this connection it was suggested in \cite{14,15} that the modified (super)conformal
transformations could be understood as a quantum deformation of the standard
(super)conformal transformations of the classical field theory.
The idea that the quantum effective action should be invariant just under the modified
(super)conformal transformations was further advanced in \cite{13}.

In the present paper we take a different viewpoint on the interplay between
the standard and modified (super)conformal transformations. We show that
any conformal field theory in $d=p+1$-dimensional Minkowski space in the phase with spontaneously
broken conformal symmetry, i.e. containing among its fields a Goldstone field (dilaton) associated with the
broken scale generator, even at the classical level can be brought, by an invertible change of variables, into
the form in which it respects invariance just under the above mentioned field-dependent
conformal transformations. This change of variables
essentially includes a field-dependent change of the Minkowski space-time coordinates
$y^\mu\; (\mu=0,1,\ldots , p)$ and maps them on the worldsheet coordinates $x^\mu$ of
the corresponding codimension-one brane in AdS$_{(d+1)}$, while the dilaton
is mapped on the brane transverse coordinate which completes $x^\mu$ to
AdS$_{(d+1)}$ in the solvable subgroup parametrization. Using this map between the conformal
and AdS bases (it can naturally be called `holographic map'),
one can rewrite any conformal field theory containing the dilaton in terms of the variables of
the corresponding AdS brane in a static gauge, and vice versa. The AdS images of the
minimal conformally invariant Lagrangians (i.e. those containing terms with no more
than two derivatives) prove to necessarily include non-minimal terms composed out of the
extrinsic curvatures of the brane. On the other hand, the conformal field theory image
of the minimal brane Nambu-Goto action is a non-polynomial and higher-derivative
extension of the minimal Minkowski space conformal actions.

In this paper we restrict our study to the bosonic case only, having in mind
to extend it to the full superconformal case in a forthcoming publication. We start with
recalling basic facts about the standard nonlinear realization of conformal group $SO(2, p+1)$ in
$p+1$-dimensional Minkowski space. Then we rewrite the algebra of $SO(2, p+1)$ in the solvable-subgroup
basis of \cite{solv} as the AdS$_{(p+2)}$ group algebra and show how to reproduce the static-gauge
Nambu-Goto action of scalar $p$-brane in AdS$_{(p+2)}$ background by applying to this group
the nonlinear realizations techniques along the line of refs. \cite{West,Iv,ik,dik}. The AdS$_{(p+2)}$ isometry
group in the second nonlinear realization acts just as the field-modified conformal transformations
of refs. \cite{5}-\cite{nscm}. Comparing two nonlinear realizations of $SO(2, p+1)$, the standard one and
the one suitable to AdS branes, we establish the explicit relation between the
coset parameters in both realizations. Finally, we give examples of various invariants in
both bases, including the conformal basis form of the Nambu-Goto action, and discuss
some possible implications of the relationship found.

\setcounter{equation}0
\section{Standard nonlinear realization of conformal \break
group in $d$ dimensions}

The algebra of the conformal group $SO(2,d)$ of $d = p+1$-dimensional Minkowski space
has the following form
\bea\label{confbasis}
&& \left[ M_{\mu\nu},M^{\rho\sigma}\right] = 2 \delta^{[\rho}_{[\mu}M_{\nu]}{}^{\sigma]}\; ,\;
\left[ P_{\mu},M_{\nu\rho}\right] =-\eta_{\mu [ \nu}P_{\rho]}\;,\;
\left[ K_{\mu},M_{\nu\rho}\right] =-\eta_{\mu [ \nu}K_{\rho]}\;, \nn
&& \left[ P_{\mu},K_{\nu}\right] =2\left( -\eta_{\mu\nu} D +2 M_{\mu\nu} \right)\;,\;
\left[ D, P_{\mu}\right] = P_{\mu}\;, \; \left[ D, K_{\mu}\right] = -K_{\mu}\;,
\eea
where
\be
A_{[\mu\nu]} \equiv \frac{1}{2}\left( A_{\mu\nu}-A_{\nu\mu} \right)\;
\ee
and $\eta_{\mu\nu} = \mbox{diag}(+-\ldots -)$. In what follows
this standard basis of conformal algebra will be called `conformal' to distinguish it from
the `AdS basis' to be specified below.

The standard nonlinear realization of the conformal group (see, e.g. \cite{cfstand}) corresponds
to choosing the Lorentz group $SO(1,p)\propto M_{\mu\nu}$ as the stability (linearization)
subgroup and so it is defined as left shifts of the following coset element
\be\label{confcoset}
g=e^{y^{\mu}P_{\mu}}e^{\Phi D} e^{\Omega^\mu K_\mu} \; .
\ee
The left shifts with parameters $a^\mu, b^\mu$ and $c$ related to the generators
$P_\mu, K_\mu$ and $D$ induce the familiar conformal transformations of the coset
coordinates
\be
\delta y^\mu = a^\mu + c\,y^\mu + 2\,(yb)y^\mu - y^2\,b^\mu~, \; \delta \Phi = c +2\,yb~,
\;\delta \Omega^\mu = e^{\Phi}\,b^\mu + 2(\Omega b)y^\mu - 2(y\Omega)\,b^\mu~. \label{trstand}
\ee

We define the left-covariant Cartan 1-forms  as follows
\bea\label{confcartan}
g^{-1}dg &=&e^{-\Phi}dy^{\mu}P_{\mu} +\left( d\Phi -2e^{-\Phi}\Omega_{\mu}dy^{\mu}\right)D-
   4e^{-\Phi}\Omega^{\mu}dy^{\nu} M_{\mu\nu}  \nn
&& + \left[ d\Omega^{\mu}-\Omega^{\mu}d\Phi +e^{-\Phi}\left(2\Omega_\nu dy^\nu \Omega^\mu -
   \Omega^2 dy^\mu\right)\right] K_\mu \; .
\eea
The vector Goldstone field $\Omega^\mu(x)$ is redundant as it can be
covariantly expressed through the only essential one, dilaton $\Phi(x)$,
by imposing the covariant Inverse Higgs constraint \cite{IH}
\be\label{confih}
\omega_D=0 \Rightarrow \Omega_\mu = {1\over 2}\,e^{\Phi} \partial_{\mu}^y \Phi~.
\ee
The remaining 1-forms associated with the coset generators then read
\be\label{confformsIH}
\omega_P^{\mu} = e^{-\Phi}dy^{\mu} \; , \; \omega_K^{\mu}=
 d\Omega^{\mu}- e^{-\Phi}\Omega^2 dy^\mu \;.
\ee
The covariant derivative of $\Omega^\mu$ is defined by the relation
\bea
&&\omega^\mu_K = \omega^\nu_P{\cal D}_\nu\Omega^\mu \quad \Rightarrow \nn
&&{\cal D}_\nu\Omega^\mu = e^{\Phi}\partial_\nu\Omega^\mu -\Omega^2\delta^\mu_\nu
= {1\over 2}
\,e^{2\Phi}\left[\partial_\nu\partial^\mu\Phi + \partial_\nu\Phi \partial^\mu \Phi
-{1\over 2}\left(\partial \Phi\partial\Phi\right)\delta^\mu_\nu \right].\label{covOm}
\eea
The covariant derivative of some non-Goldstone (`matter') field $\Psi^a(y)$, where $a$ is
an index of the Lorentz group representation, is defined by
\bea
&& d\Psi^a -4e^{-\Phi}\Omega^\mu dy^\nu(M_{\mu\nu})^a_b\Psi^b =
\omega^\mu_P{\cal D}_\mu \Psi^a\quad \Rightarrow \nn
&&{\cal D}_\mu\Psi^a = e^\Phi\,\partial_\mu\Psi^a +4\Omega^\nu(M_{\mu\nu})^a_b\Psi^b. \label{covPsi1}
\eea
When $y^\mu$ is transformed according to \p{trstand}, the field $\Psi^a$, as well as the covariant
derivatives \p{covOm} and \p{covPsi1}, undergo an induced Lorentz rotation with respect
to their Lorentz indices, e.g.,
\be
\delta \Psi^a(y) =\Psi^a{}'(y')- \Psi^a(y) = \beta^{\mu\nu}(M_{\mu\nu})^a_b\Psi^b(y)~, \;\;
\beta^{\mu\nu} = -4\,y^{[\mu}b^{\nu]}~. \label{genTr1}
\ee

The conformally invariant measure of integration over $\{y^\mu \}$ is defined as
the exterior product of $d$ 1-forms $\omega^\mu_P$
\be
S_1 = \int \mu(y) = \int d^{(p+1)} y\, e^{-(p+1) \Phi}~.\label{S1}
\ee
It can be treated as the conformally invariant potential of dilaton.

The covariant kinetic term of $\Phi$ can be constructed as
\be
S_\Phi^{kin} = \int d^{(p+1)} y\, e^{-(p+1) \Phi}\,{\cal D}_\mu\Omega^\mu =
{1\over 4}(p-1)\int d^{(p+1)}y
\, e^{(1-p)\Phi}\,\partial \Phi\partial\Phi \label{Skin1}
\ee
(while passing to the final form of \p{Skin1}, we integrated by parts).
For the special case $d=2\, (p =1)$ the Lagrangian in \p{Skin1} is reduced to a full derivative.
In this case one can still define the non-tensor kinetic term which is invariant
under \p{trstand} up to a shift by full derivative
\be
S_\Phi^{kin(2)} = {1\over 2}\int d^2y\, \partial\Phi \partial \Phi~. \label{kin21}
\ee

Conformally invariant Lagrangians of matter fields $\Psi^a$ are obtained by
replacing ordinary derivatives by the covariant ones \p{covPsi1} and promoting $d^{(p+1)} y$
to the conformally invariant measure \p{S1}. E.g., the standard Maxwell field strength
can be covariantized as
\be
\tilde{F}_{\mu\nu} =  {\cal D}_\mu \tilde{A}_\nu -  {\cal D}_\nu \tilde{A}_\mu =
e^{2\Phi}F_{\mu\nu}~, \label{Fconf}
\ee
where $\tilde{A}_\mu$  is transformed according to the generic law \p{genTr1} and its
covariant derivative is defined by \p{covPsi1}. It is related in the following way
to the ordinary Maxwell vector potential $A_\mu$ having the same conformal
transformation law as the partial derivative $\partial_\mu$ and the standard gauge
transformation law
\be
\tilde{A}_\mu = e^{-\Phi}A_\mu~, \quad F_{\mu\nu} = \partial_\mu A_\nu - \partial_\nu A_\mu~.
\ee
The conformally invariant action of $A_\mu$ then reads
\be
S^{(c)}_{M} = -{1\over 4}\int d^{(p+1)}y\, e^{(3-p)\Phi} F^{\mu\nu}F_{\mu\nu}~. \label{Maxconf1}
\ee
At $d=4\;(p=3)$ it coincides with the standard Maxwell action which is conformal
in its own right only in this dimension.

This formalism of nonlinear realizations of conformal symmetry is universal
in the following sense. In any theory in which conformal symmetry
is spontaneously broken, it is always possible to make a field redefinition which splits
the full set of scalar fields of the theory into the dilaton $\Phi$
with the transformation law \p{trstand} and the subset of fields which are scalars of
weight zero under conformal transformations. For instance, let us consider
the free action of $N$ massless scalar fields $\phi^I$, $I =1,\ldots N$ ($p\neq 1$):
\be
S = \int d^{(p+1)}y\; \partial\phi^I\partial \phi^I~. \label{SphiI}
\ee
It is invariant under \p{trstand} (up to a shift of the Lagrangian by a full derivative)
if $\phi^I$ are transformed with the appropriate weight
\be
\delta \phi^I = {1\over 2}\,(1-p)(c + 2yb)\phi^I~, \quad \delta |\phi| = {1\over 2}\,(1-p)(c +yb)|\phi|~.
\ee
If some field develops a non-zero vacuum value, $<\phi^{I_0}> = v\neq 0$ (e.g. due to the presence
of some conformally invariant potential term which should be added to \p{SphiI}),
the conformal symmetry is spontaneously broken and one can perform the
equivalence field redefinition
\bea
&& \phi^I = \frac{|\phi|}{v}\,\hat{\phi}^I~, \quad \hat{\phi}^I \hat{\phi}^I = v^2~,\quad \delta \hat{\phi}^I
= 0 \label{algconstr}\\
&& |\phi| = v + \tilde{\phi} + \ldots = v\,e^{{1\over 2}(1-p)\Phi}~, \quad \phi^{I_0} \equiv \tilde{\phi} + v~.
\label{redef22}
\eea
Then the action \p{SphiI}, up to an overall coefficient and surface terms, can be rewritten as
\be
S = \int d^d y e^{(1-p)\Phi}\left[{1\over 4} (1-p)^2\partial \Phi \partial \Phi
+ \partial \hat{\phi}^I \partial \hat{\phi}^I\right].
\ee
The first term coincides with the universal dilaton action \p{Skin1} while the second term is
the action of a nonlinear sigma model of the internal symmetry group realized
on the indices $I$.

An example of the system admitting such a field redefinition is supplied, e.g.,
by the scalar fields sector of ${\cal N}=4, d=4$ SYM action in the Coulomb branch. Consider, e.g. the simplest case
of $SU(2)$ gauge group. When some scalar field
valued in the Cartan subalgebra $u(1)$ acquires a non-zero
expectation value (which is a solution of classical equations of motion for the full action including
the conformally invariant quartic potential of the scalar fields),
the gauge group gets broken to $U(1)$
and there remain $6$ scalar massless fields in the theory which form a vector
of the $R$-symmetry group $SO(6) \sim SU(4)$. The norm of this vector is just
the dilaton associated with the spontaneous breaking of conformal symmetry $SO(2,4)$.
The remaining 5 independent fields appear as the solution of the algebraic constraint in
\p{algconstr} and parametrize the internal sphere $S^5 \sim SO(6)/SO(5)$. Thus  the
set of 6 massless bosonic fields of $SU(2)$ $N=4$ SYM theory in the Coulomb branch
naturally splits into the $SO(6)$ invariant dilaton sector and
the sector of a nonlinear sigma model on $S^5$.

In the special case of $d=2\;(p=1)$ the field $\phi^I$ is a
scalar of the conformal weight zero, so no redefinition like \p{algconstr}, \p{redef22}
is needed. The kinetic and potential terms of dilaton \p{kin21}, \p{S1} can be
independently added, if necessary. An example of such $d=2$ system, which, like
${\cal N}=4$ SYM is conformal (and superconformal) both on classical and quantum
levels, is provided by ${\cal N} = (4,4)$ supersymmetric $SU(2)$ WZW sigma model \cite{ik2}.
Its bosonic sector includes four scalar fields, one of which is a dilaton and
three remaining ones possess zero conformal weight and parametrize
the coset $S^3 \sim SU(2)\times SU(2)/SU(2)$. The conformally
invariant bosonic action is a sum of free action of the dilaton and standard
$SU(2)$ WZW action \cite{wit1}.

\setcounter{equation}0
\section{The AdS nonlinear realization}
In the AdS basis we introduce the following generators
\be\label{adsgenerators}
\hK_{\mu} =mK_{\mu}-\frac{1}{2m}P_{\mu}\;,\; \hD=mD \;,
\ee
where $m$ will be identified with the inverse radius of AdS space.

The same conformal algebra \p{confbasis} in the AdS basis \p{adsgenerators}
reads
\bea\label{adsbasis}
&& \left[ \hK_{\mu},\hK_{\nu}\right] =-4 M_{\mu\nu}\; ,\;
\left[ P_{\mu},\hK_{\nu}\right] =2\left( -\eta_{\mu\nu} \hD +2m M_{\mu\nu} \right),\nn
&&\left[ \hD, P_{\mu}\right] = mP_{\mu}\;, \; \left[ \hD, \hK_{\mu}\right] = -\left(P_{\mu} +m\hK_{\mu}\right)
\eea
(commutators with the Lorentz generators $M_{\mu\nu}$ are of the same form as in \p{confbasis}).

The basic difference of \p{adsbasis} from \p{confbasis} is that the generators
$(\hat{K}^\mu, M_{\mu\nu})$ generate the semi-simple subgroup $SO(1,d)$ of $SO(2,d)$,
while the subgroup $(K^\mu, M_{\mu\nu})$ has the structure of a semi-direct product.
As a result, in the coset element \p{confcoset} rewritten in the new basis
\be\label{adscoset}
g=e^{x^{\mu}P_{\mu}}e^{q \hD} e^{\Lambda^\mu \hK_\mu} \; ,
\ee
the coordinates $x^\mu$ and $q(x)$ are parameters of the coset manifold $SO(2,d)/SO(1,d)$
which is none other than AdS$_{(d+1)}$. This parametrization of AdS$_{(d+1)}$ was
called in \cite{solv} `the solvable subgroup parametrization', since the generators $P_\mu$
and $\hat{D}$ with which the AdS$_{(d+1)}$ coordinates are associated as the coset parameters
constitute the maximal solvable subgroup of $SO(2,d)$. One more convenience of the basis \p{adsbasis}
with the manifestly included dimensionful parameter $m$ is that one can perform
the contraction $m =0$ in \p{adsbasis}, which takes it just into the $(d+1)$-dimensional
Poincar\'e group $ISO(1,d)$, with the set $(P_\mu, \hat{D})$ becoming the generators of
$(d+1)$-translations. In this limit $x^\mu$ and ${1\over \sqrt{2}}q$ are recognized as
the coordinates of $(d+1)$-dimensional
Minkowski space, the standard $R = \infty$ limiting case of AdS$_{(d+1)}$. This confirms
the interpretation of the parameter $m$ as the inverse AdS$_{(d+1)}$ radius.

In the new basis the Cartan forms \p{confcartan} read
\bea\label{adscartan}
g^{-1}dg &=& \left[ e^{-mq}\left( dx^{\mu}+\frac{\lambda^\mu \lambda_\nu dx^\nu}{1-\frac{\lambda^2}{2}}
  \right) -\frac{\lambda^\mu dq}{1-\frac{\lambda^2}{2}}\right] P_\mu \nn
  && +\, \frac{1+\frac{\lambda^2}{2}}{1-\frac{\lambda^2}{2}}\left[ dq- 2\frac{e^{-mq}\lambda_\mu dx^{\mu}}
    {1+\frac{\lambda^2}{2}}\right] \hD \nn
  && +\, \frac{1}{1-\frac{\lambda^2}{2} }\left[ d\lambda^\mu -m\lambda^\mu dq -me^{-mq}\left(
    \lambda^2dx^\mu -2 \lambda^\mu\lambda_\nu dx^\nu\right)\right] \hK_\mu \nn
  && +\, \omega^{\mu\nu}_M M_{\mu\nu}~,
\eea
where
\be
\lambda^\mu = \frac{\mbox{tanh} \sqrt{\frac{\Lambda^2}{2}}}{\sqrt{\frac{\Lambda^2}{2}}}\Lambda^\mu \;.
\ee
and the new basis form of $\omega^{\mu\nu}_M = -4\,e^{-\Phi}\Omega^{[\mu}y^{\nu]}$ can be found
using the explicit relation between the parameters of the coset elements \p{confcoset} and
\p{adscoset} which will be given in the next Section.

The inverse Higgs constraint \p{confih} is rewritten in the AdS basis as
\bea\label{adsIH}
&& \omega_{\hat{D}} = 0 \quad
\Rightarrow \quad \frac{\lambda_\mu}{1+\frac{\lambda^2}{2}}= {1\over 2}e^{m q}\,\partial_\mu q~, \nn
&& \lambda_\mu =
e^{mq}\,\frac{\partial_\mu q}{1 + \sqrt{1 - {1\over 2}e^{2mq}(\partial q\partial q)}}~.
\eea
On the surface of this covariant constraint the remaining coset space Cartan forms
are given by the expressions:
\bea \label{adsformsIH}
&& \omega_P^\mu= e^{-mq}\left( \delta^\mu_\nu -
\frac{\lambda^\mu \lambda_\nu}{1+\frac{\lambda^2}{2}}\right)dx^\nu \equiv E^\mu_\nu dx^\nu
= e^{-mq}\hat{E}^\mu_\nu dx^\nu\;, \nn
&& \omega_{\hat K}^\mu=\frac{1}{1-\frac{\lambda^2}{2}}\left( d\lambda^\mu - m \lambda^2 \omega_P^\mu \right).
\eea

The covariant derivative, with the Lorentz connection part omitted, is defined by
\be
dx^\mu \partial_\mu = \omega^\mu_P {\cal D}_\mu \Rightarrow {\cal D}_\mu = e^{mq}\left( \delta_\mu^\nu +
  \frac{\lambda_\mu \lambda^\nu}{1-\frac{\lambda^2}{2}}\right)\partial_\nu \equiv
(E^{-1})_\mu^\nu\partial_\nu = e^{mq}(\hat{E}^{-1})_\mu^\nu \partial_\nu\;. \label{defDads}
\ee
The covariant derivative of the $SO(1,d+1)/SO(1,d)$ Goldstone field $\lambda^\mu$ is
defined by the formula analogous to \p{covOm}
\bea
&& \omega^\nu_{\hat K}=\omega^\mu_P \cD_\mu \lambda^\nu~, \nn
&& \cD_\mu \lambda^\nu=\frac{1}{1-\frac{\lambda^2}{2}}\left[ e^{mq}\left(\delta_\mu^\rho
+ \frac{\lambda_\mu \lambda^\rho}{1-\frac{\lambda^2}{2}}\right)\partial_\rho\lambda^\nu -
  m\lambda^2 \delta_\mu^\nu \right].\label{covLam}
\eea

It is straightforward to find the transformation laws of $x^\mu, q(x)$ and $\lambda^\mu(x)$
under the left shifts of \p{adscoset}
\bea
&&\delta x^\mu = a^\mu + c\,x^\mu + 2\,(xb)x^\mu - x^2\,b^\mu + {1\over 2m^2}\,e^{2mq} b^\mu~, \;
\delta q = {1\over m}(c +2\,xb)~, \label{modconf} \\
&&\delta \lambda^\mu = {1\over m}\left(1+\frac{\lambda^2}{2}\right)e^{mq}\,\hat{E}^\mu_\nu\, b^\nu
+ 2(\lambda b)x^\mu - 2(x\lambda)\,b^\mu~, \label{modconf1}
\eea
where all group parameters are the same as in \p{trstand}. It is easy to check that \p{modconf}
are perfectly consistent with the inverse Higgs expression \p{adsIH} for $\lambda^\mu(x)$.

The transformations of $x^\mu$ and $q(x)$ are just the field-dependent conformal transformations which were
discussed in \cite{mald}, \cite{5}-\cite{nscm} in connection with the AdS branes
and were shown in \cite{solv} to naturally arise
as the AdS isometries in the above solvable-subgroup parametrization of AdS groups. To see how this
interpretation is recovered in the present approach, let us first write the AdS$_{(d+1)}$ metric
\be
ds^2 = \omega_P^\mu\omega_{P\mu} = e^{-2mq}dx^\mu\eta_{\mu\nu}dx^\nu - dq^2~. \label{dist1}
\ee
The change of variables (we assume $p \neq 1$)
\be
e^{-2mq} = \left(\frac{U}{R} \right)^{4\over p-1}\frac{2}{(p-1)^2}~, \quad R = {1\over m}~,\label{change}
\ee
brings \p{dist1} (up to a factor) and transformation rules \p{modconf} into the form
\bea
&& ds^2 =  \left(\frac{U}{R} \right)^{4\over p-1}dx^\mu\eta_{\mu\nu}dx^\nu  -
\left(\frac{R}{U} \right)^{2} dU^2~, \label{dist2} \\
&&\delta x^\mu = a^\mu + c\,x^\mu + 2\,(xb)x^\mu - x^2\,b^\mu +
{1\over 4}(p-1)^2\,\frac{R^{2\frac{p+1}{p-1}}}{U^{\frac{4}{p-1}}}\, b^\mu~,\nn
&& \delta U = -{1\over 2}(p-1)(c +2 xb)U~, \label{modconf2}
\eea
which coincide with those given e.g. in \cite{5} (up to a rescaling of $x^\mu$ and a different
choice of the signature of Minkowski metric).

The simplest invariant of the nonlinear realization considered is again the covariant volume of $x$-space
obtained as the integral of wedge product of $(p+1)$ 1-forms $\omega_P^\mu$. The difference from \p{S1}
is that this invariant is basically the static-gauge Nambu-Goto (NG) action for $p$-brane in AdS$_{(p+2)}$
\bea\label{NGaction}
S_{NG} &=& - \int d^{(p+1)} x \left[ \mbox{det}\,E - e^{-(p+1)mq}\right]=
 \int d^{(p+1)} x\, e^{-(p+1)mq} \left(1 - \frac{1-\frac{\lambda^2}{2}}{1+\frac{\lambda^2}{2}}\right) \nn
&=& -\int d^{(p+1)} x\, e^{-(p+1)mq} \left[\sqrt{1 - {1\over 2}e^{2mq}(\partial q\partial q)}-1\right],
\eea
where we used the relations
\be
\mbox{det}\,E = e^{-(p+1)mq}\,\mbox{det}\,\hat{E}~,\quad \mbox{det}\,\hat{E} =
\frac{1-{\lambda^2\over 2}}{1+{\lambda^2\over 2}} = \sqrt{1 - {1\over 2}e^{2mq}(\partial q\partial q)}
\label{det}
\ee
and subtracted 1 to obey the standard requirement of absence of the vacuum energy (corresponding to
$q = const$) \cite{mald}. Note that the subtracted term
\be
S_2 = \int d^{(p+1)} x\, e^{-(p+1)mq}\label{S2}
\ee
is invariant under \p{modconf} (up to a shift of the integrand by a full derivative) on its own. In
most interesting cases it is a part of some WZ (or CS) term in a static gauge. The action \p{NGaction}
is universal, in the sense that it describes the radial (pure AdS) part of any AdS$_n\times S^m$
$(n+m-2)$-brane action corresponding to `freezing' (setting equal to constants) all other fields
on the brane (e.g., the gauge fields and angular $S^5$ fields in the case of AdS$_5\times S^5$ D3-brane)
and also to neglecting some further possible WZ-type terms on the brane worldvolume. Actually, this
universality extends to the branes on AdS$_n\times X^m$ where $X^m$ can stand for some $m$-dimensional curved
manifold different from the sphere, e.g. one of the manifolds considered in \cite{klwit} while analysing
the AdS/CFT correspondence for a general ${\cal N}=4$ SYM theory in the Coulomb branch.

The minimal covariant actions of various `matter' fields are obtained via replacing the ordinary derivatives
by the covariant ones and inserting $\mbox{det}\,E$ into the integration  measure. E.g., the covariant
kinetic term of some scalar field $\phi(x)$ is given by
\be
S_\phi = \int d^{(p+1)}x\,\mbox{det}\hat{E}\, e^{(p-1)mq}\,
\hat{G}^{\mu\nu}\partial_\mu\phi\partial_\nu\phi~, \label{Sphi}
\ee
where
\be
\hat{G}^{\mu\nu} = \eta^{\omega\rho}(\hat{E}^{-1})^\mu_\omega (\hat{E}^{-1})^\nu_\rho =
\eta^{\mu\nu} + e^{2mq}\,\frac{2}{1 - {1\over 2}e^{2mq}(\partial q\partial q)}\,\partial^\mu q
\partial^\nu q \label{invG}
\ee
is the inverse of the induced metric
\be
\hat{G}_{\mu\nu} = \eta_{\omega\rho}E^\omega_\mu E_\nu^\rho = \eta_{\mu\nu} - {1\over 2}e^{2mq}
\partial_\mu q \partial_\nu q \label{dirG}
\ee
(with the factors $e^{\pm 2mq}$ detached).

As the last topic of this Section, let us clarify the geometric meaning of the covariant derivative
\p{covLam} which plays an important role in our construction. We will show that it is the tangent-space
projection of the first extrinsic curvature of the brane. For simplicity, we shall consider the
limiting case $m=0$ in \p{covLam} and \p{adsIH} which corresponds to the $p$ brane in the flat $(p+2)$-dimensional
Minkowski background. The generalization to the AdS case is straightforward.

One defines the extrinsic curvature by the relation (see, e.g.\cite{pol}-\cite{extr2})
\be
\nabla_\mu\partial_\nu X^An_A = K_{\mu\nu}~, \label{extrcur}
\ee
where $X^A$ are target brane coordinates, $X^A = (x^\mu, -{1\over \sqrt{2}}q)$ in the considered
static gauge, $\eta_{AB} = (\eta_{\mu\nu}, -1)$, $n_A = (n_\mu, n)$ is a normal to the brane worldsheet
\be
\partial_\mu X^A\,n_A =0~, \quad n^An_A = n^\mu n_\mu - n^2 = -1 \label{ort}
\ee
and
\be
\nabla_\mu\partial_\nu X^A = (\partial_\mu\partial_\nu - \Gamma^\rho_{\mu\nu}\partial_\rho)X^A~.
\ee
The induced metric $G_{\mu\nu}$ in the static gauge and its inverse $G^{\mu\nu}$ are given by \p{dirG}, \p{invG}
with $m=0$. We find
\be
\Gamma^\rho_{\mu\nu}= G^{\rho\omega}\Gamma_{\mu\nu\;\omega}~, \quad
\Gamma_{\mu\nu\;\omega} = {1\over 2}\left(\partial_\mu G_{\nu\omega} + \partial_\nu G_{\mu\omega} -
\partial_\omega G_{\mu\nu} \right) = -{1\over 2}\partial_\mu\partial_\nu q \partial_\omega q~,
\ee
and
\be
\nabla_\mu\partial_\nu q = \frac{1}{1 -{1\over 2}(\partial q\partial q)}\,\partial_\mu\partial_\nu q\;,\quad
\nabla_\mu\partial_\nu x^\rho =
{1\over 2}\frac{1}{1-{1\over 2}(\partial q\partial q)}\,\partial_{\mu}\partial_\nu q\partial^\rho q~.
\ee
Further, the orthogonality condition \p{ort} in the static gauge is reduced to \footnote{Actually, this condition
is another form of the inverse Higgs constraint \p{adsIH} at $m=0$, with $n_\mu$ being related via a field redefinition
to the Goldstone field $\lambda_\mu$.}
\be
n_\mu + {1\over \sqrt{2}}\partial_\mu q\,\sqrt{1 + n^\nu n_\nu} =0 \quad \Rightarrow \quad
n_\mu = -{1\over \sqrt{2}} \,\frac{\partial_\mu q}{\sqrt{1-{1\over 2}(\partial q\partial q)}}~.
\ee
After substituting all this into the definition \p{extrcur}, we obtain
\be
K_{\mu\nu} = {1\over \sqrt{2}}\frac{1}{\sqrt{1-{1\over 2}(\partial q\partial q)}}\,\partial_\mu\partial_\nu q
\ee
and
\be
{\cal D}_\mu \lambda_\nu = {1\over \sqrt{2}}(E^{-1})_\mu^\rho (E^{-1})_\nu^\omega K_{\rho\omega}~.
\ee

\setcounter{equation}0
\section{An equivalence relation between CFT and AdS bases}
In both nonlinear realizations described above we deal with the same coset manifold $SO(2,d)/SO(1,d-1)$,
in which the coset parameters are divided into the space-time coordinates and Goldstone
fields in two different ways. In the first realization the coordinates $y^\mu$ parametrize the $d$-dimensional Minkowski
space considered as a coset of $SO(2,d)$ identified with the corresponding conformal
group.\footnote{To be more rigorous, it is the compactified Minkowski space which can be treated as
a coset manifold of conformal group.} All other parameters are Goldstone fields, the essential one being
dilaton $\Phi(y)$ associated with the spontaneous breaking of scale invariance. In the second realization
the space-time coordinates $x^\mu$ on their own do not constitute a coset manifold of $SO(2,d)$ and
therefore do not form a closed set under the left action of this group. However, together with the Goldstone field
$q(x)$ they parametrize the coset $SO(2,d)/SO(1,d) \sim$ AdS$_{(d+1)}$ and this extended set is closed
under the action of $SO(2,d)$. These coset parameters admit a clear interpretation as the worldvolume ($x^\mu$)
and transverse ($q$) coordinates of $(d-1)$-brane evolving in AdS$_{(d+1)}$.

Apart from this essential difference in the interpretation, the fact that both these realizations
(with vector Goldstone fields $\Omega_\mu$ and $\lambda_\mu$ included) are
in fact defined on the same full coset of $SO(2,d)$, viz. $SO(2,d)/SO(1,d-1)$, suggests the existence
of relation between these two different coset parametrizations. This relation can be straightforwardly
extracted from comparison of \p{confcoset} and \p{adscoset}
\be
y^\mu=x^\mu-\frac{e^{mq}}{2m}\lambda^\mu \;, \; \Phi=mq+\ln\left(1-\frac{\lambda^2}{2}\right)\; , \;
\Omega^\mu=m\lambda^\mu \; . \label{map}
\ee
We see that it is invertible at any finite non-zero $m = 1/R$. It is straightforward to check that
the Minkowski space conformal transformations \p{trstand} are mapped by \p{map} on the field-dependent ones
\p{modconf} and vice versa. Since this change of variables maps the geometric objects living in the
AdS$_{(d+1)}$ bulk on those defined on its Minkowski boundary, it seems natural to name it
`holographic transformation'. It is important to emphasize that this holographic transformation
essentially involves the Goldstone field $\lambda^\mu$ (or $\Omega_\mu$) which basically becomes the
derivative of $q(x)$ (or $\Phi(y)$) after imposing the
covariant constraint \p{adsIH} (or its conformal basis counterpart \p{confih}). However, for the existence of
map \p{map} it does not matter
whether \p{adsIH} or \p{confih} are imposed or not, the only necessary
condition is the presence of vector parameters $\Omega^\mu(y)$ and $\lambda^\mu(x)$ in
both cosets. In other words, \p{map} could not be guessed solely in the framework of the
pure AdS$_{(d+1)}$ geometry, i.e. by dealing with the AdS coordinates $x^\mu$
and $q$ alone; it can be defined only when considering extended coset manifolds
$\{y^\mu,\Phi, \Omega_\mu\}$ and $\{x^\mu, q, \lambda_\mu\}$. Another characteristic feature
of the map \p{map} is that it is well defined only for non-zero and finite values of AdS radius $R = 1/m$.

Using the holographic transformation, any conformal field theory in Minkowski space
with a dilaton among its basic fields can be projected onto the variables of AdS brane and vice versa.
To find the precise form of various $SO(2,d)$ invariants in two bases, the conformal and AdS ones,
let us first define the transition matrix
\be
\frac{\partial y^\nu}{\partial x^\mu}\equiv
  {\cal A}_\mu^\nu=\delta_\mu^\nu -\frac{\lambda_\mu \lambda^\nu}{1+\frac{\lambda^2}{2}} -
   \frac{e^{mq}}{2m}\partial_\mu \lambda^\nu = \left(1-\frac{\lambda^2}{2}\right)\hat{E}_\mu^\rho\, T^\nu_\rho\;,
\label{defA}
\ee
where
\be
T^\nu_\rho = \delta^\nu_\rho - {1\over 2m}{\cal D}_\rho\lambda^\nu~,\label{defT}
\ee
the matrix $\hat{E}^\mu_\nu$ is defined by \p{adsformsIH} and ${\cal D}_\rho \lambda^\nu$ is
the covariant derivative of $\lambda^\nu$ defined
in \p{covLam} (it is an extrinsic curvature of the brane). We then
have the following general formula for the Jacobian of the change of space-time
coordinates in \p{map}
\be
J \equiv \mbox{det}\,{\cal A} = \left(1-\frac{\lambda^2}{2}\right)^{p+1}\mbox{det}\,\hat{E}\,
\mbox{det}\,T~. \label{Jac}
\ee

Making the change of variables \p{map} in the invariant dilaton Lagrangians \p{S1} and \p{Skin1},
we obtain, respectively,
\bea
S_1 &=& \int d^{(p+1)}y\,e^{-(p+1)\Phi} = \int d^{(p+1)}x\,e^{-(p+1)m q}\ \mbox{det}\, \hat{E}\, \mbox{det}\, T
 \nn
&=& \int d^{(p+1)}x\,e^{-(p+1)m q}\,\sqrt{1 -{1\over 2}e^{2mq}(\partial q \partial q)}\, \mbox{det}\, T~,
\label{S1transf} \\
S_\Phi^{kin} &=& \int d^{(p+1)}y\,e^{-(p+1)\Phi}\ {\cal D}_\mu\Omega^\mu =
{1\over 2}\int d^{(p+1)}y\,e^{(1-p)\Phi}\left[\Box \Phi +{1\over 2}(1-p)(\partial\Phi\partial\Phi)\right] \nn
&=& m\,\int d^{(p+1)}x\,e^{-m(p+1)q}\ \mbox{det}\hat{E}\,\left[\mbox{det} T\,(T^{-1}{\cal D}\lambda)^\mu_\mu \right]\nn
&=& m\,\int d^{(p+1)}x\,e^{-m(p+1)q}\,\sqrt{1 -{1\over 2}e^{2mq}(\partial q \partial q)}\,
\left[\mbox{det} T\,(T^{-1}{\cal D}\lambda)^\mu_\mu \right].
\label{Skintransf}
\eea

We observe a surprising fact that the AdS image of the potential term of dilaton contains
the NG part of the AdS $p$-brane action \p{NGaction} modified by the higher-derivative covariants
collected in $\mbox{det} (I - {1\over 2m}{\cal D}\lambda) = 1 -{1\over 2m}{\cal D}_\mu\lambda^\mu +\ldots\;$.
As we saw, ${\cal D}_\mu\lambda^\nu$ is basically the extrinsic curvature of the $p$-brane.
So already the simplest conformal invariant in Minkowski space proves to produce, on the AdS side,
a rather complicated action which is the standard $p$-brane action in AdS$_{(p+2)}$ plus corrections composed
out of the extrinsic curvature tensor. The leading (with two derivatives) term in the r.h.s.
of \p{S1transf} comes both from the NG square root and the terms $\sim \partial_\mu\lambda^\mu\,,\;\lambda^2$
in ${\cal D}_\mu\lambda^\mu$ (see \p{covLam} and \p{adsIH})
\be
S_1 = \int d^4 x \,e^{-(p+1)mq}\left[1 - {1\over 8}(p+1)e^{2mq}(\partial q\partial q) + \ldots \right].
\ee
Note that in the flat case $m=0$ the extrinsic curvature terms are capable to produce
only higher-order (in fields and derivatives) corrections to the minimal
NG $p$-brane action (as follows from the expression \p{covLam} at $m=0$).
On the other hand, the AdS image of the kinetic term
of dilaton, eq. \p{Skintransf}, starts with the correct kinetic term of $q$:
\be
S^{kin}_\Phi = \frac{m^2}{4}\int d^4 x \left[e^{-(p-1)mq}(p-1)\,(\partial q\partial q) + \ldots \right].
\ee
Note, however, that it comes solely from the extrinsic curvature term, not from the NG square root.
The latter is always multiplied by degrees of the extrinsic curvature in \p{Skintransf}.

A way to elude this paradox of generating kinetic terms from the pure potential ones
via the change of variables could be to start from the reasonable field theory action
on the CFT side, having from the beginning both kinetic and potential dilaton terms,
i.e. from the action
\be
S = S_\Phi^{kin} + \gamma S_1~, \label{sumact}
\ee
where $\gamma$ is a coupling constant. To the second order in $\partial_\mu q$ it is
\be
S =  \int d^4 x \,\left(\gamma\, e^{-(p+1)mq} +{1\over 4}[m^2(p-1)- {1\over 2}\gamma (p+1)]
(\partial q\partial q) + \ldots \right),
\ee
and we observe that the holographic transformation \p{map} merely renormalizes the
coefficient before the kinetic term. Nevertheless, the paradox still persists because
one can fully eliminate the kinetic term of $q$ by choosing $\gamma = 2m^2\,\frac{p-1}{p+1}\;$.
Then on the CFT side we still have quite reasonable field theory, while on the AdS side
we get an action admitting no standard weak-field expansion. These observations suggest
that the map \p{map} is not the standard equivalence transformation preserving the canonical
structure of the given theory. This peculiarity of \p{map} is manifested, first, in that the essential
part of \p{map} is a non-linear field-dependent transformation of the space-time coordinate starting
with a derivative of $q$ and, second, in that the relation between $\Phi$ and $q$ contains
a shift by kinetic term of $q$, $\Phi = mq - {1\over 8}(\partial q\partial q) + \ldots\;$.
Note that for the conformal actions containing no potential terms of
dilaton the relations \p{map} can be
still treated as setting a genuine equivalence map, since they always take the kinetic term of $\Phi$
into that of $q$ (up to rescaling by $m$) plus some terms of higher order in $q$
and its derivatives. The same remains true when bringing the minimal AdS brane
action \p{NGaction} with vanishing vacuum energy into the conformal basis (see next Section).

In the special $d=2\,(p=1)$ case the conformally invariant kinetic term of $\Phi$ is given by
the non-tensor Lagrangian \p{kin21}. Its AdS image is also of non-tensor form, in
contradistinction to the manifestly invariant term \p{Skintransf} for $d\neq 2$
\bea
S_{\Phi}^{kin(2)}&=&\int d^2 y\, (\partial \Phi\partial \Phi) = 4m^2 \int d^2x
  \frac{e^{-2mq}\,\lambda^2}{\left(1-\frac{\lambda^2}{2}\right)^2}\,\mbox{det}\, {\cal A} \nn
&=&
4m^2 \int d^2x\,e^{-2mq}\,\lambda^2\,\mbox{det}\, \hat{E}\,\mbox{det}\, T~. \label{kin22}
\eea
It is not easy to check the invariance of \p{kin22} under the transformations \p{modconf}.
For proving that \p{kin22} is indeed invariant, up to a shift of the Lagrangian by a full derivative,
one needs to use the explicit form of $\mbox{det}\,{\cal  A}$ for this case
\bea
\mbox{det}\, {\cal A} &=& \frac{1}{2}\left[ \left(\mbox{Tr}\,{\cal A}\right)^2-\mbox{Tr}\,{\cal A}^2)\right]  =
\frac{1-\frac{\lambda^2}{2}}{1+\frac{\lambda^2}{2}}\left[
   1-\frac{e^{mq} \partial_\mu\lambda^\mu}{2m}-\frac{e^{mq}\lambda^\mu\lambda^\nu\partial_\mu\lambda_\nu}
     {2m\left( 1-\frac{\lambda^2}{2} \right)}\right] \nn
&& +\, \frac{e^{2mq}}{8m^2}\left[ \left( \partial_\mu\lambda^\mu\right)^2 -
     \partial_\mu\lambda^\nu\partial_\nu\lambda^\mu \right].
\eea

The AdS images of the conformally invariant kinetic terms of `matter' fields can be
obtained by making the variable change \p{map} in the corresponding actions. For instance,
for a scalar field $\Psi(y)$ we find
\bea
&& S_{\psi} = \int d^{(p+1)}y\,e^{(p-1)\Phi}\,\partial^\mu \Psi\partial_\mu \Psi
= \int d^{(p+1)}x\,\mbox{det}\, E\, {\cal L}(q, \Psi)~, \nn
&& {\cal L}(q,\Psi) = \mbox{det}\, T\ \eta^{\mu\nu}(T^{-1})^\omega_\mu(T^{-1})^\tau_\nu\ {\cal D}_\omega\Psi
{\cal D}_\tau\Psi = G^{\mu\nu}\partial_\mu\Psi\partial_\nu\Psi + O({\cal D}\lambda)~, \label{matADS} \\
&& {\cal D}_\mu\Psi = (E^{-1})_\mu^\nu\partial_\nu\Psi~, \quad G^{\mu\nu} =
\eta^{\rho\tau}(E^{-1})_\rho^\mu (E^{-1})_\tau^\nu~. \nonumber
\eea
We see that this expression differs from the minimal covariantization \p{Sphi} by
couplings to the brane extrinsic curvatures.

The change \p{map} brings the conformal Maxwell action \p{Maxconf1} into the form
\be
S_M = -{1\over 4}\int d^{(p+1)}x\,\mbox{det}\, E\,{\cal H}^{\mu\nu}{\cal H}_{\mu\nu}~, \label{maxads1}
\ee
where
\bea
&& {\cal H}_{\mu\nu} = (T^{-1})_\mu^\rho(T^{-1})_\nu^\omega {\cal F}_{\rho\omega}~, \;
{\cal F}_{\mu\nu} = (E^{-1})_\mu^\rho(E^{-1})_\nu^\omega \hat{F}_{\rho\omega}~, \nn
&& \hat{F}_{\rho\omega} = \partial^x_\rho\hat{A}_\omega - \partial^x_\omega\hat{A}_\rho~, \;
\hat{A}_\mu = {\cal A}_\mu^\nu A_\nu~. \label{Fads}
\eea
Once again, a difference from the minimal invariant Lagrangian
$\sim {\cal F}^{\mu\nu}{\cal F}_{\mu\nu}$ $=$
$G^{\mu\nu}G^{\omega\lambda}$ $\hat{F}_{\mu\omega}\hat{F}_{\nu\lambda}$ is the presence
of extra couplings with the extrinsic curvature.

It is instructive to give how $\hat{A}_\nu$ and $\hat{F}_{\mu\nu}$ are transformed under
\p{modconf}. Their transformation laws follow from the property that $A_\mu$ is transformed under
the conformal group as the derivative $\partial^y_\mu$, while the matrix ${\cal A}^\mu_\nu =
\partial y^\mu/\partial x^\nu$ as
\be
\delta {\cal A}^\mu_\nu = 2(yb - xb){\cal A}^\mu_\nu + 2{\cal A}_\nu^\rho(b_\rho y^\mu -y_\rho b^\mu)
 - 2\left(b_\nu x^\rho - x_\nu b^\rho + {1\over 4m^2}\partial_\nu e^{2mq}\,b^\rho\right){\cal A}_\rho^\mu~.
\label{Atransf}
\ee
Then
\be
\delta \hat{A}_\mu = -(c + 2xb)\hat{A}_\mu -
2\left(b_\mu x^\rho - x_\mu b^\rho + {1\over 4m^2}\partial_\mu e^{2mq}\,b^\rho\right)\hat{A}_\rho~,\label{Aconftr}
\ee
or
\be
\delta^* \hat{A}_\mu = \hat{A}_\mu'(x) - \hat{A}_\mu(x) = \delta^*_c\hat{A}_\mu -{1\over 2m^2}e^{2mq}\,b^\rho\hat{F}_{\rho\mu}
-{1\over 2m^2}\partial_\mu\left(e^{2mq}b^\rho\hat{A}_\rho\right),\label{Aconftr2}
\ee
where $\delta^*_c$ denotes the conventional conformal (including no $q$-dependent terms) part of the complete
variation. The transformation of ${\hat F}_{\mu\nu}$ is of standard form
$$
\delta \hat{F}_{\mu\nu} = -(\partial_\mu \delta x^\rho)\, \hat{F}_{\rho\nu} -
(\partial_\nu \delta x^\rho)\,\hat{F}_{\mu\rho}~.
$$

\setcounter{equation}{0}
\section{AdS brane actions in the conformal basis}
In the previous section we have found how the simplest conformally invariant Lagrangians in Minkowski
space look after passing to the AdS basis. It is of interest also to see what the AdS brane action
\p{NGaction} looks like in the conformal basis, with the conventionally realized spontaneously
broken conformal symmetry. The helpful relations are
\be
{\cal D}_\mu\Omega^\nu =m(T^{-1})_\mu^\omega {\cal D}_\omega\lambda^\nu~, \quad
(T^{-1})_\mu^\nu = \delta^\nu_\mu + {1\over 2m^2}{\cal D}_\mu\Omega^\nu~,
\ee
where ${\cal D}_\mu\Omega^\nu$ was defined in \p{covOm}.

We start with the `potential' term of $q$, eq. \p{S2}. Making in \p{S2} the change of variables
inverse to \p{map}, we find
\be
S_2 = \int d^{(p+1)}y\, e^{-(p+1)\Phi}\ \frac{1+ {1\over 8m^2}\,e^{2\Phi}\,
(\partial \Phi\partial \Phi)}{1 -{1\over 8m^2}\,e^{2\Phi}\,
(\partial \Phi\partial \Phi)}\,\mbox{det}\left(I +{1\over 2m^2}{\cal D}\Omega \right)~. \label{S2conf}
\ee
For the pure NG-part of the action \p{NGaction} we obtain rather simple expression
\be
S = \int d^{(p+1)}y\,e^{-(p+1)\Phi}\,\mbox{det}\left(I + {1\over 2 m^2}{\cal D}\Omega \right).
\label{NG1}
\ee
Then the full brane action \p{NGaction} takes the form
\be
S_{NG} = {1\over 4m^2}\int d^{(p+1)}y\,e^{(1-p)\Phi}\,
\frac{(\partial \Phi\partial\Phi)}{1-{1\over 8m^2}e^{2\Phi}(\partial \Phi\partial \Phi)}\,
\mbox{det}\left(I + {1\over 2 m^2}{\cal D}\Omega \right).\label{NGconf}
\ee
Thus we have found an equivalent representation of the static-gauge action \p{NGaction}
of $p$-brane in AdS$_{(p+2)}$ as a non-linear extension of the conformally-invariant
dilaton action in $(p+1)$ dimensional Minkowski space. Note that the conformal image
of the brane action is nonlinear and non-polynomial, however it is a rational
function of $\Phi$ and its derivatives. We also note that, despite the simplicity of
the standard conformal transformations \p{trstand}, it is rather tricky to directly check
that \p{NGconf} or \p{S2conf} are indeed invariant under them. The difficulty
originates from the property that the Lagrangian densities in \p{NGconf},
\p{S2conf}, like their AdS images \p{NGaction}, \p{S2}, are not tensors, they are shifted by a full
derivative under \p{trstand} (as distinct from the Lagrangian in \p{NG1} which is
manifestly invariant). Though the conformal variation of $S_{NG}$ \p{NGconf} can easily
be found
\be
\delta_c S_{NG} = {1\over m^2}\int d^{(p+1)}y\,e^{(1-p)\Phi}\,
\frac{b^\mu\,\partial_\mu\Phi}{\left[1-{1\over 8m^2}e^{2\Phi}(\partial \Phi\partial \Phi)\right]^2}\,
\mbox{det}\left(I + {1\over 2 m^2}{\cal D}\Omega \right), \label{NGconfvar}
\ee
it is far from obvious that the integrand in \p{NGconfvar} is a full derivative.
To see this, one should demonstrate that the variational
derivative of \p{NGconfvar} is identically vanishing,
$$
\frac{\delta}{\delta \Phi(y)} (\delta_c S_{NG})= 0~.
$$
The proof makes use of the explicit expressions \p{covOm} and \p{confih} and is somewhat
tiresome, though straightforward. Notice the crucial importance of terms with
two derivatives on $\Phi$ coming from the determinant in \p{NGconfvar}. As a simpler exercise,
one can directly check that \p{NGconfvar} is reduced to a full derivative
in the first order in $1/m^2$ (since transformations \p{trstand} do not include $m^2$,
each term in the expansion of \p{NGconf}
in powers of $1/m^2$ should be invariant separately). It would hardly be possible to guess such
a non-tensor conformal invariant, staying solely in the framework of the standard nonlinear
realization of conformal group.

Our last example will be the conformal field theory image of the full bosonic part of
D3-brane on AdS$_5\times S^5$. Neglecting the `magnetic' part of the Chern-Simons term,
the action in the static gauge can be written as
(see, e.g. \cite{revTs})
\be
S_5 = -C\int d^4x\,\frac{|X|^4}{R^4} \left[\sqrt{-\mbox{det}\left(\eta_{\mu\nu}  -\frac{R^4}{|X|^4}\,
\partial_\mu X^i\partial_\nu X^i + \frac{R^2}{|X|^2}\,\hat{F}_{\mu\nu}\right)}\, -1\right],\label{ads5}
\ee
where $i=1,\ldots 6$, $|X| = \sqrt{X^iX^i}$, $C$ is some positive renormalization constant the
precise form of which is of no interest in the present context and the signs are adjusted in accordance with our
choice of the Minkowski metric $\eta_{\mu\nu} = \mbox{diag}\,(+---)$.

Firstly we rewrite  \p{ads5} in our notation, using the field redefinition
\be
\frac{R}{|X|} = {1\over \sqrt{2}}\,e^{mq}~, \quad m= {1\over R}~,
\ee
which is the particular $p=3$ case of the redefinition \p{change}. We obtain
\be
S_5 = -4C\int d^4x\,e^{-4mq}\left[(\mbox{det}\,\hat{E})\;\sqrt{-\mbox{det}\left(\eta_{\mu\nu} + {1\over 2}{\cal F}_{\mu\nu}
-{1\over 2}{\cal D}_\mu\tilde{X}^i{\cal D}_\nu\tilde{X}^i \right)} -1\right],\label{ads51}
\ee
where ${\cal D}_\mu$ and ${\cal F}_{\mu\nu}$ were defined in \p{defDads}, \p{Fads} and $\tilde{X}^i$ parametrize
the sphere $S^5$,
$$
\tilde{X}^i\tilde{X}^i = R^2~.
$$
For constant $\tilde{X}^i$ and $\hat{A}_\mu$ the action \p{ads51} is reduced
to the pure AdS$_{(d+1)}$ action \p{NGaction} with $d=4$.

Now, making in \p{ads51} the change of variables inverse to \p{map}, we obtain the conformal basis form of
the AdS$_5\times S^5$ action
\bea
S_5 &=& 4C\int d^4y\,e^{-4\Phi}\,\mbox{det}\left(I + {1\over 2 m^2}{\cal D}\Omega \right)
\left\{\frac{1+ {1\over 8m^2}\,e^{2\Phi}\,
(\partial \Phi\partial \Phi)}{1 -{1\over 8m^2}\,e^{2\Phi}(\partial \Phi\partial \Phi)} \right. \nn
&& \left. -\,\sqrt{-\mbox{det}\left[\eta_{\mu\nu} + {1\over 2}\,e^{2\Phi}
T_\mu^\rho\,T^\omega_\nu \left(F_{\rho\omega}
-\partial_\rho\tilde{Y}^i \partial_\omega\tilde{Y}^i \right)\right]}\right\},\label{ads52}
\eea
where
\be
\tilde{Y}^i(y)\equiv \tilde{X}^i(x(y)) = \frac{R}{|Y|}Y^i~, \; \frac{R}{|Y|} = {1\over \sqrt{2}}\,e^\Phi\,
\frac{1}{1 -{1\over 8m^2}e^{2\Phi}(\partial \Phi \partial \Phi)}~.\label{ads53}
\ee
Thus we have succeeded in equivalently rewriting the effective bosonic action of D3-brane
in the AdS$_5\times S^5$ background \p{ads5} or \p{ads51} as a conformally invariant
nonlinear action of the coupled system of the following set of fields in 4-dimensional Minkowski
space $\{y^\mu\}$: dilaton $\Phi(y)$, five independent scalar fields
$\tilde{Y}^i(y)\,, \;\tilde{Y}^i\tilde{Y}^i = R^2\,$, parametrizing
the sphere $S^5$, and an abelian gauge field $A_\mu(y)$. For $\tilde{Y}^i$ and $A_\mu$
we still have a version of the Dirac-Born-Infeld action promoted to a conformally-invariant one
due to couplings to the dilaton $\Phi(y)$. It also includes extra conformal couplings to the
curvature ${\cal D}_\mu\Omega^\nu$ (through the common factor
$\mbox{det}\left(I + {1\over 2 m^2}{\cal D}\Omega \right)$
and the matrices $T^\rho_\mu$ in the determinant under the square root). The dilaton $\Phi(y)$ itself,
with all other fields neglected, is described
by the nonlinear higher-derivative action \p{NGconf}. The crucial difference between \p{ads5}
(or \p{ads51}) and \p{ads52} is that the latter involves fields having standard transformation properties
under the conformal group $SO(2,4)$, while in \p{ads5} the latter is realized as the group of isometry of
AdS$_5$, with transformations depending on $|X|$. The group $SO(6)$ has the same realization
in both representations as the isometry group of 5-sphere $S^5$.

\setcounter{equation}{0}
\section{Discussion}
In this paper we have found a new kind of holographic relation between field theories possessing spontaneously broken
conformal symmetry in $d$-dimensional Minkowski space and the codimension-$(n+1)$ branes in AdS$_{(d+1)}\times X^n$
type backgrounds in the static gauge (with the sphere $S^n$ as a particular case of $X^n$).
This relation takes place already at the classical level and transforms
the dilaton Goldstone field associated with the spontaneous breaking of scale invariance into the
transverse (or radial) brane co-ordinate completing the $d$-dimensional brane worldvolume to
the full AdS$_{(d+1)}$ manifold. It does not touch the $X^n$-valued part of transverse coordinates which
are described by a kind of nonlinear sigma model action in both representations. The conformally invariant
minimal actions in Minkowski space including the dilaton are transformed into the highly nonlinear actions
given on the AdS brane worldvolume and involving, as their essential part, couplings to the extrinsic curvature
of the brane. Conversely, the standard worldvolume AdS brane effective actions prove to be equivalent to some
non-polynomial conformally invariant actions in the Minkowski space. This map is one-to-one (at least, classically)
for the conformal actions containing no dilaton potential and for brane actions with the vanishing
vacuum energy. The geometric origin of this map can be revealed most clearly within the
nonlinear realization description of AdS branes \cite{dik} which generalizes the analogous description of
branes in the flat backgrounds \cite{West,Iv,ik}. In particular, it turns out that the standard realization of
the conformal group in the Minkowski space and its transverse brane coordinate-dependent realization
as the AdS$_{(d+1)}$ isometry group in the solvable-subgroup parametrization of AdS$_{(d+1)}$
are simply two alternative ways of presenting symmetry of the same system.

As the most interesting subjects for further study we mention the generalization of the above
relationship to the case of AdS superbranes and, respectively, superconformal symmetries,
as well as the understanding of how it can be promoted to the quantum case.

Since the appropriate framework for the
bosonic case is provided by nonlinear realizations of conformal groups, we expect that
the generalization to the supersymmetry case can be fulfilled most naturally within the
PBGS (Partial Breaking of Global Supersymmetry) approach to superbranes (see \cite{pbgs} and
refs. therein). In the given context the PBGS approach amounts to describing AdS superbranes
in terms of superfield nonlinear realizations of the appropriate superconformal group, with
half of supersymmetries (special conformal supersymmetries) being nonlinearly realized and
the rest providing manifest linear invariances of the corresponding actions. The superanalog of
the map \p{map} should relate different coset superspaces of superconformal groups: those
where these groups are realized in the standard way, i.e. with the superspace coordinates
transforming through themselves without any mixing with the Goldstone superfields (see, e.g.
\cite{ferb,scfstand}),
and those where the transformation laws of superspace coordinates essentially involve
the Goldstone superfields, like the modified bosonic transformations \p{modconf}. The
second type of realizations should be relevant to the PBGS superbrane actions with superextensions of
AdS$\times S$ manifolds as the target supermanifolds for which the appropriate superconformal
groups define superisometries. An example of the worldvolume superfield PBGS action for AdS superbranes,
that of the AdS$_4$ supermembrane, was recently constructed in \cite{dik}. The relevant
Goldstone superfield-dependent realization of the corresponding superisometry group $OSp(1|4)$
($N=1,\, d=3$ superconformal group) on the $N=1, d=3$ worldvolume superspace coordinates was
explicitly found.

As for generalizing the map \p{map} to the quantum case, one should firstly understand how
to treat the field dependence of the change of space-time coordinates in \p{map} in this case.
Since the fields $q$ and $\Phi$ will not longer commute with their derivatives, it seems
that the transformed coordinates should also be non-commuting. To keep \p{map} invertible,
for consistency one should require both coordinate sets $\{y^\mu\}$ and $\{x^\nu\}$ to be
non-commuting. This could provide a link with the non-commutative geometry.

We shall finish with a few further comments on possible implications of
the holographic map \p{map}.

In the AdS/CFT context the actions of standard conformal field theories are usually treated
as a the $R \rightarrow 0$ (or low-velocity) approximation of the AdS brane
effective worldvolume actions. For instance, the $U(1)$ part of the ${\cal N}=4$ $SU(2)$ SYM
action in the Coulomb branch can be recovered as the $R \rightarrow 0$ limit
of the abelian D3-brane action on AdS$_5\times S^5$. Indeed, for the bosonic part of the
latter, eq. \p{ads5}, we have
$$
S_5 \sim \int d^4x \left[{1\over 2}\partial^\mu X^i\partial_\mu X^i
- {1\over 4}\hat{F}^{\mu\nu}\hat{F}_{\mu\nu} + O(R)\right]~.
$$
In this limit the field-dependent conformal transformations \p{modconf}, \p{modconf2}
are reduced to the standard ones which are characteristic of the field theory actions (in \p{modconf}
one needs to rescale $q \rightarrow R q$ to approach this limit in an unambiguous way).

The existence of the holographic map \p{map} suggests a different view of the relationship between
the conformal field theory actions and the worldvolume actions of AdS superbranes.
As we saw, any conformal field theory action in the branch
with spontaneously broken conformal symmetry, after singling out the dilaton field, can be
rewritten in terms of the AdS brane variables, with the field-modified conformal transformations
defining the relevant symmetry. This relationship exists at any finite and non-vanishing AdS radius
$R= 1/m\,$. We observed, however, that the AdS images of conformal field theory dilaton actions
do not coincide with the standard NG type brane actions, but are given by the expressions of the
type \p{S1transf}, \p{Skintransf} which essentially include powers of extrinsic curvature
of the brane.\footnote{An interesting
exception \cite{ikn} is the $d=1$ case of conformal mechanics where \p{Skintransf} coincides,
up to a full derivative,
with the $d=1$ case of \p{NGaction}.} Besides, the AdS
images of other fields do not appear under the square root as, e.g. in the
standard AdS$_5\times S^5$ D3-brane action \p{ads5}, but have the form \p{matADS}, \p{maxads1}
where all nonlinearities are solely due to the AdS brane transverse coordinate $q(x)$
and its derivatives. It is interesting to further explore
this surprising `brane' representation of (super)conformal field theories, especially in the quantum
domain, and to better understand the role of couplings to extrinsic curvature which are unavoidable
in this representation. In this connection, let us recall that a string
with `rigidity', i.e. with extrinsic curvature terms added to the action, was considered as
a candidate for the QCD string \cite{pol} (see also \cite{extr1,extr2}). We also notice that the
higher-derivative corrections to the minimal worldvolume superbrane actions are
$\kappa$-invariant extensions of the extrinsic curvature terms (see \cite{holi} and refs. therein).

Besides addressing the obvious problem of studying AdS$_5\times S^5$ brane representation of the full
${\cal N}=4,\, d=4$ SYM action (both in the component and superfield approaches),
it would be instructive to investigate analogous representations of the actions of some
superconformal theories in lower dimensions, e.g. the action of ${\cal N}=(4,4), \,d=2$ WZW
sigma model \cite{ik2} which was mentioned in the end of Sect. 2. Since its bosonic sector
in the standard (conformal) basis includes the dilaton and the $S^3 \sim SU(2)\times SU(2)/SU(2)$
coset fields, it should admit a representation in terms of variables of
superstring on AdS$_3\times S^3\,$.

One more possible implication of the holographic AdS/CFT map is as follows.
As was already mentioned, the worldvolume action of some probe superbrane in
the AdS$_n\times S^m$ type background (obtained as a solution of the appropriate supergravity)
is expected to be recovered on the CFT side as a sum of the leading (and subleading) terms
in the loop expansion of the low-energy quantum effective action of the related (super)conformal
field theory taken in a phase with spontaneously broken (super)conformal symmetry \cite{chep,mald2,mald}.
If the quantum field theory is arranged to respect non-anomalous rigid symmetries of the classical
theory, it is reasonable to assume that there exists a formulation of its quantum effective action
(e.g., in the appropriate background field formalism) such that it is still invariant under
the standard conformal group. Then for checking the above mentioned `supergravity-CFT'
correspondence one is led to compare the quantum effective action just with
the conformal basis form of the corresponding superbrane
worldvolume action, i.e. with expressions like \p{NGconf}, \p{ads52}. In the context of
the correspondence between the Coulomb branch of ${\cal N}=4$ SYM and
abelian D3-branes on AdS$_5\times S^5$
this reasoning implies that the scalar field sector
of the ${\cal N}=4$ SYM quantum effective action should be of the form \p{ads52} rather
than \p{ads5} or \p{ads51}. The latter expressions are to be recovered only after performing
the holographic transformation \p{map}.
As a rule, the correspondence discussed is checked for the gauge field sector only,
by setting scalar fields to be constants \cite{ats}. From \p{ads52} and \p{ads53} it is seen that
in this approximation $\Phi = mq$, and \p{ads52} actually coincides with \p{ads51} or \p{ads5}. It would
be of interest to explore the structure of the scalar field sector of the low-energy $N=4$ SYM effective
action beyond this constant field approximation and compare it with \p{ads52}.\footnote{One should
restore the omitted `magnetic' 5-form Chern-Simons term in \p{ads52} while checking this.}

\section*{Acknowledgments}
We are grateful to Paolo Pasti,
Dmitri Sorokin and Mario Tonin for useful discussions.
This work was partially supported by the Fondo Affari Internazionali
Convenzione Particellare INFN-JINR, the European Community's
Human Potential Programme under contract HPRN-CT-2000-00131 Quantum Spacetime,
INTAS grant No. 00-00254, grant
DFG 436 RUS 113/669 as well as RFBR-CNRS grant No. 01-02-22005.

\end{document}